\documentclass[10pt,twocolumn]{article}          % LaTeX 2e 
\usepackage{latex8}                              % LaTeX 2e 
\usepackage{times}                               % LaTeX 2e 
\usepackage{psfig}

\begin{document}

\title{A Scientific Data Management System for Irregular
Applications\thanks{This work was supported in part by the
Mathematical, Information, and Computational Sciences Division
subprogram of the Office of Advanced Scientific Computing Research,
U.S.\ Department of Energy, under Contract W-31-109-Eng-38, and in part
by a Work-for-Others Subaward No.\ 751 with the University of
Illinois, under NSF Cooperative Agreement \#ACI-9619019.}}

\author{
Jaechun~No$^\dag$ ~~ Rajeev~Thakur$^\dag$ ~~ Dinesh~Kaushik$^\dag$ ~~
Lori~Freitag$^\dag$ ~~ Alok~Choudhary$^\ddag$ \\ \\
\begin{tabular}{c c}
\hskip .05in \dag Math.\ and Computer Science Division &
 \hskip .05in \ddag Dept.\ of Elec. and Computer Eng. \\
\hskip .05in Argonne National Laboratory & \hskip .05in Northwestern
University\\
\hskip .05in Argonne, IL 60439 & \hskip .05in Evanston, IL 60208\\
{\tt \{jano,thakur,kaushik,freitag\}@mcs.anl.gov} & \hskip .05in {\tt choudhar@ece.nwu.edu}
\end{tabular}
}

\maketitle
\thispagestyle{empty}

\begin{abstract}
Many scientific applications are I/O intensive and generate large data
sets, spanning hundreds or thousands of ``files.'' Management,
storage, efficient access, and analysis of this data present an
extremely challenging task. We have developed a software system, called
Scientific Data Manager (SDM), that uses a combination of parallel
file I/O and database support for high-performance scientific data
management. SDM provides a high-level API to the user and, internally,
uses a parallel file system to store real data and a database to store
application-related metadata.
In this paper, we describe how we designed and implemented SDM to
support irregular applications. SDM can efficiently handle the reading
and writing of data in an irregular mesh, as well as the distribution of
index values. We describe the SDM user interface and how we have
implemented it to achieve high performance. SDM makes extensive use of
MPI-IO's noncontiguous collective I/O functions.  SDM also uses the
concept of a {\em history file} to optimize the cost of the index
distribution using the metadata stored in database. We present
performance results with two irregular applications, a CFD code called
FUN3D and a Rayleigh-Taylor instability code, on the SGI Origin2000
at Argonne National Laboratory.
\end{abstract}

%------------------------------------------------------------------------- 
\Section{Introduction \label{sec:intro}}
Many large-scale scientific applications are I/O intensive and
generate large amounts of data (on the order of several hundred
gigabytes to terabytes)~\cite{delrosario:prospects,poole:sio-survey}.
Many of these applications perform their computation and I/O on an
irregularly discretized mesh.  The data accesses in those applications
make extensive use of arrays, called indirection
array~\cite{D94,ravi95a} or map array~\cite{grop99a}, in which each
value of the array denotes the corresponding data position in memory
or in the file.

The data distribution in irregular applications can be done either
by using compiler directives with the support of runtime
preprocessing~\cite{rein:fortran,HPFF93} or by using a runtime
library~\cite{D94,ravi95a}.  Most of the previous work in the area of
unstructured-grid applications focuses mainly on computation and
communication in such applications, not on I/O.

We have developed a software system for large-scale scientific data
management, called Scientific Data Manager (SDM)~\cite{jano:sc00}, 
that combines the good features of both file I/O and
databases. SDM provides a high-level, user-friendly
interface. Internally, SDM interacts with a database to store
application-related metadata and uses MPI-IO to store the real data on
a high-performance parallel file system. SDM takes advantage of
various I/O optimizations available in MPI-IO, such as collective I/O
and noncontiguous requests, in a manner that is transparent to the
user.  As a result, users can access data with the performance of
parallel file I/O, without having to bother with the details of file
I/O.

In a previous paper~\cite{jano:sc00}, we described the use of SDM for
regular applications. In this paper, we describe the API, design, and
implementation of SDM for irregular applications. SDM can efficiently
handle the reading and writing of data in an irregular mesh, as well as
the distribution of index values. SDM also uses the concept of a {\em
history file} to optimize the cost of the index distribution using the
metadata stored in database. We present performance results with two
irregular applications, a CFD code called FUN3D and a Rayleigh-Taylor
instability code, on the SGI Origin2000 at Argonne National
Laboratory.

The rest of this paper is organized as follows. In
Section~\ref{sec:obj} we discuss our goals in developing SDM for
irregular problems.  In Section~\ref{sec:impl} we present a typical
irregular problem and describe the detailed implementation issues of
SDM to solve the problem.  Performance results on the SGI Origin2000
at Argonne National Laboratory are presented in
Section~\ref{sec:perf}. We discuss related work in
Section~\ref{sec:related} and conclude in Section~\ref{sec:conc}.

\Section{Design Objectives \label{sec:obj}}

Our main objectives in designing SDM for irregular applications were
to achieve high-performance parallel I/O, to provide 
a convenient high-level API,
and to optimize the execution cost of irregular applications.

\begin{itemize}
\item{\bf High-Performance I/O}.
To achieve high-performance I/O, we decided to use a parallel file-I/O
system to store real data and use MPI-IO to access this data.  MPI-IO,
the I/O interface defined as part of the MPI-2
standard~\cite{grop99a,mpi97a}, is rapidly emerging as the standard,
portable API for I/O in parallel applications. MPI-IO is specifically
designed to enable the optimizations that are critical for
high-performance parallel I/O. Examples of these optimizations include
collective I/O, the ability to access noncontiguous data sets, and the
ability to pass hints to the implementation about access patterns,
file-striping parameters, and so forth.

\item{\bf High-Level API}.
Our goal was to provide a high-level unified API for any kind of
application (regular or irregular) while encapsulating the details of
either MPI-IO or databases.  With SDM, user can specify the data with a
high-level description, together with annotations, and use a similar
API for data retrieval.  SDM internally translates the user's request
into appropriate MPI-IO calls, including creating MPI derived
datatypes for noncontiguous data~\cite{thak98a}.  SDM also interacts
with the database when necessary, by using embedded SQL functions.

\item{\bf Optimization for Irregular Applications}.
In irregular applications, the cost of an index distribution is
usually expensive, in terms of communication and computation.
In SDM, after partitioning the index values among processes, the local
index subsets of all processes are asynchronously written to a {\em
history file}, and the associated metadata is stored in database.  When
the same index distribution is needed in subsequent runs, the index
values are read from the history file using the metadata stored
in database, and thereby the user can avoid repeating the
communication and computation for the same index distribution.
\end{itemize}

\Section{Implementation \label{sec:impl}}

We discuss the SDM API for solving a sample irregular problem and show
 how the API is implemented.

\SubSection{An Irregular Problem and SDM API}

\begin{figure}[h]
\centerline{\psfig{figure=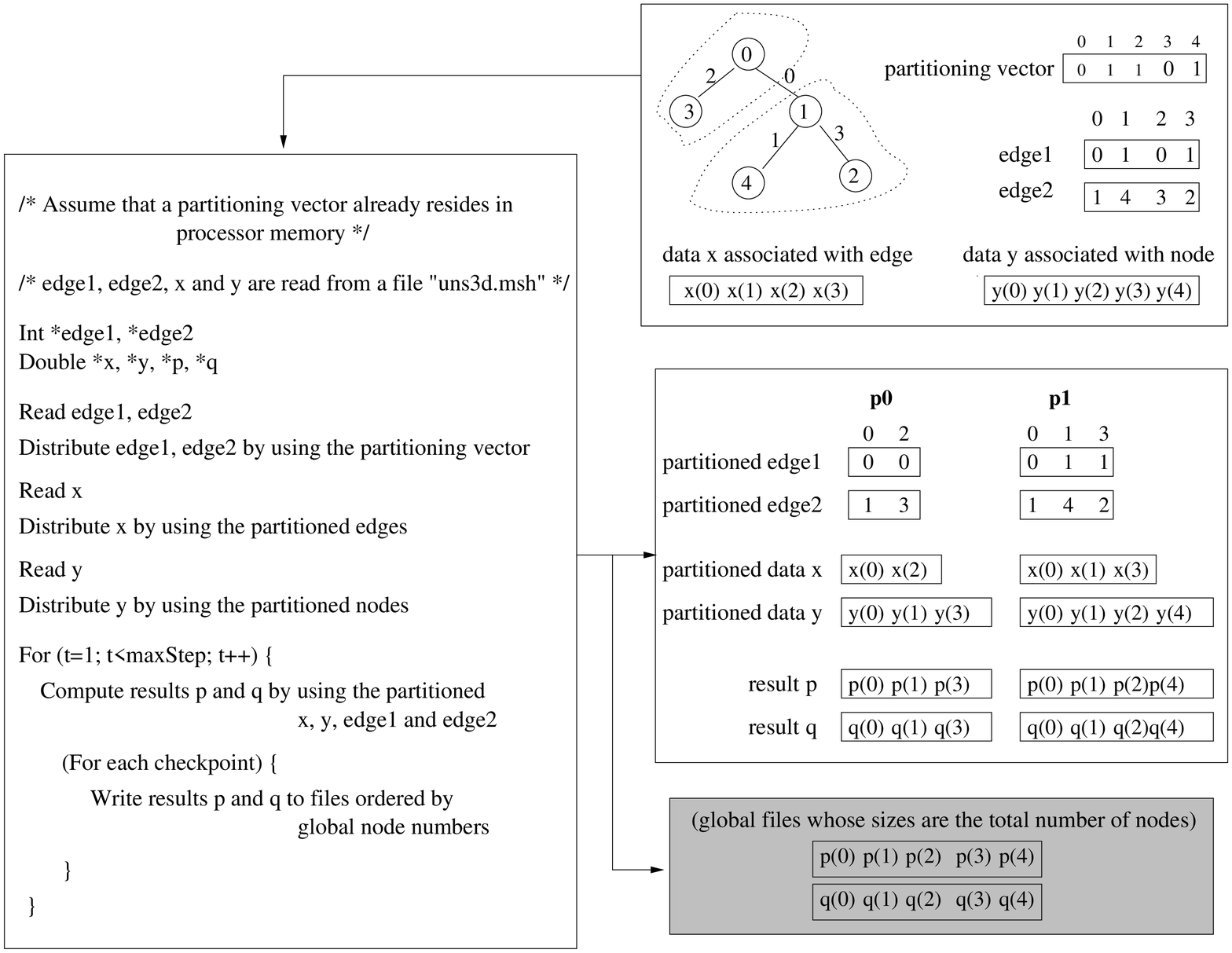,height=3.5in,width=3.5in}}
\caption{A sample irregular problem and its solution}
\label{fig:problem}
\end{figure}

Figure~\ref{fig:problem} shows a typical irregular problem that sweeps
over the edges of an irregular mesh.  In this problem, {\tt edge1} and
{\tt edge2} are two arrays representing nodes connected by an edge,
and arrays {\tt x} and {\tt y} are the actual data associated with
each edge and node, respectively.  The partitioned arrays of {\tt
edge1}, {\tt edge2}, {\tt x}, and {\tt y} contain a single level of
``ghost data'' beyond the boundaries to minimize remote accesses.
After the computation is completed, the results {\tt p} and {\tt q}
are written to a file in the order of global node numbers.

Figures~\ref{fig:API1} and~\ref{fig:API2} respectively show the SDM API for
writing the results {\tt p} and {\tt q} and for partitioning {\tt
edge1}, {\tt edge2}, {\tt x}, and {\tt y} among processes to solve the
problem described in Figure~\ref{fig:problem}.  We use the term {\em
import} to distinguish it from a {\em read} operation.  A read
operation reads the data created in SDM, whereas an import operation
reads the data created outside of SDM.

\begin{figure}[h]
\fbox{
\begin{minipage}[t]{3.2in}
\hspace*{0.1in}SDM\_initialize(nameOfApplication); \\
\hspace*{0.1in}result = SDM\_make\_datalist(2, \{p, q\}); \\
\hspace*{0.1in}result[0].data\_type = DOUBLE; \\
\hspace*{0.1in}SDM\_associate\_attributes(2, \&result[0]); \\
\hspace*{0.1in}handle = SDM\_set\_attributes(2, result); \\
\hspace*{0.1in}...... \\
\hspace*{0.1in}/* Partition edge1, edge2, x and y among processes \\
\hspace*{0.5in} (Figure 3) */ \\
\hspace*{0.1in}...... \\
\hspace*{0.1in}SDM\_data\_view(handle, 2, p, \&vector, \&localNodes); \\
\hspace*{0.1in}For (t=1; \(t<maxStep\); t++) \{ \\
\hspace*{0.3in}...... \\
\hspace*{0.3in}Do computation and produce results p and q; \\
\hspace*{0.3in}...... \\
\hspace*{0.3in}For (each checkpoint) \{ \\
\hspace*{0.5in}SDM\_write(handle, p, t, pBuf); \\
\hspace*{0.5in}SDM\_write(handle, q, t, qbuf); \\
\hspace*{0.3in}\} \\
\hspace*{0.1in}\} \\
\hspace*{0.1in}SDM\_finalize(handle, 2); \\
\end{minipage}
}
\caption{SDM API for writing results}
\label{fig:API1}
\end{figure}

\SubSection{Implementation Details}

The {\em partitioning vector}
 is the one generated from a partitioning tool, 
 such as MeTis\cite{george:metis,kirk:metis}. Each value
of the vector denotes a processor rank where the node should be assigned.
In SDM, the partitioning vector should be replicated among processes.
Next, the {\em map array} is the one that specifies the mapping
of each element of the local array to the global array. This map array
is created in SDM after partitioning
the indexes using a partitioning vector,
or the map array can be specified by the user.

Figure~\ref{fig:API1} shows the steps involved in initializing SDM to solve
the problem in Figure~\ref{fig:problem}.
Running the problem on SDM begins by calling
the {\em SDM\_initialize} to
establish database connection (for storing metadata). 
Six database tables,
{\em run\_table}, {\em access\_pattern\_table}, {\em execution\_table},
{\em import\_table}, {\em index\_table}, and {\em index\_history\_table},
are created to store the metadata associated with the application.
Since two
data sets, {\tt p} and {\tt q}, are produced as a result of
computations and they have the same data type and global size,
these data sets are grouped in a data group to experiment different ways
of organizing data in files. All the metadata associated with these data sets
are stored in a database in the {\em SDM\_set\_attributes}.

Figure~\ref{fig:API2} describes the steps in SDM
to partition the indexes and data.
The four arrays, {\tt edge1}, {\tt edge2}, {\tt x}, and {\tt y}, are
imported by creating a data group. Since these arrays
have been created outside of SDM, the user has no control 
over the arrays except to read them, by specifying their data type,
appropriate file offset, and length. The user need not
create several data groups to import the arrays.
In the {\em SDM\_make\_importlist}, the metadata of this
imported data group, including a mechanism for the import
(partition), is stored in the {\em import\_table} for a later use.

In order to partition {\tt edge1} and {\tt edge2}, the {\em SDM\_import}
is called to import the arrays with the parameters of file handle,
their position in the data group, file offset,
file length, and user buffer to hold
the data. 
\begin{figure}[h]
\fbox{
\begin{minipage}[t]{3.2in}
\hspace*{0.1in}import = SDM\_make\_datalist(4, \{edge1, edge2, x, y\}); \\
\hspace*{0.1in}import[2].data\_type = DOUBLE; \\
\hspace*{0.1in}SDM\_associate\_attributes(2, \&import[2]); \\
\hspace*{0.1in}SDM\_make\_importlist(handle, 4, import); \\
\hspace*{0.1in} \\
\hspace*{0.1in}SDM\_import(handle, edge1, 0, totalEdges, tmp); \\
\hspace*{0.1in}SDM\_import(handle, edge2, (totalEdges*sizeof(int)), \\
\hspace*{0.3in}totalEdges, tmp+(totalEdges*sizeof(int))); \\
\hspace*{0.1in} \\
\hspace*{0.1in}/* Distribute edge1 and edge2 among processes */ \\
\hspace*{0.1in}vector = SDM\_partition\_table(handle, \\
\hspace*{0.5in} partitioning\_vector, totalNodes); \\
\hspace*{0.1in}partitioned\_edge = SDM\_partition\_index(handle, \\
\hspace*{0.3in}partitioning\_vector, totalNodes, \&tmp, \&vector); \\
\hspace*{0.1in} \\
\hspace*{0.1in}localEdges = SDM\_partition\_index\_size(handle); \\
\hspace*{0.1in}localNodes = SDM\_partition\_data\_size(handle); \\
\hspace*{0.1in} \\
\hspace*{0.1in}/* Make a history of this index distribution */ \\
\hspace*{0.1in}SDM\_index\_registry(handle, partitioned\_edge, vector); \\
\hspace*{0.1in} \\
\hspace*{0.1in}/* Import x */ \\
\hspace*{0.1in}file\_offset = 2*totalEdges*sizeof(int); \\
\hspace*{0.1in}SDM\_data\_view(handle, 1, x, \&partitioned\_edge, \\
\hspace*{0.3in}\&localEdges); \\
\hspace*{0.1in}SDM\_import(handle, x, file\_offset, totalEdges, xBuf); \\ 
\hspace*{0.1in} \\
\hspace*{0.1in}/* Import y */ \\
\hspace*{0.1in}file\_offset += totalEdges * sizeof(double); \\
\hspace*{0.1in}SDM\_data\_view(handle, 1, y, \&vector, \&localNodes); \\
\hspace*{0.1in}SDM\_import(handle, y, file\_offset, totalNodes, yBuf); \\
\hspace*{0.1in} \\
\hspace*{0.1in}SDM\_release\_importlist(handle, 4); \\
\end{minipage}
}
\caption{SDM API for partitioning indexes and data}
\label{fig:API2}
\end{figure}
The {\em SDM\_import} first accesses the
 {\em index\_table} in the database
to see whether a history file exists with this problem size.
If so, the metadata, such as each process's partitioned index size and
the history file name, is retrieved from the {\em index\_table} and
{\em index\_history\_table}, and the control exits the {\em SDM\_import}.
Otherwise, the desired data is imported to the application.
Since {\tt edge1} and {\tt edge2} are being imported in a
contiguous way, there is no need to specify data mapping between the file
and processor memory. In the {\em SDM\_import},
the total domain (file length) is equally divided among processes, and
the data in the domain is contiguously imported into the application.
In our example, edges {\tt 0} and {\tt 1} are imported to process 0, and
edges {\tt 2} and {\tt 3} are imported to process 1.

In the {\em SDM\_partition\_table},
the global partitioning vector,
 {\tt partitioning\_vector} in Figure~\ref{fig:API2}, is
converted to the local vector, {\tt vector} in Figure~\ref{fig:API2},
to determine
which node should be assigned
to which process. In the example, nodes {\tt 0} and {\tt 3} are assigned
to process 0, and nodes {\tt 1}, {\tt 2}, and {\tt 4}
are assigned to process 1.

If there is a history file for this problem size,
the {\em SDM\_partition\_index}
reads the already partitioned {\tt edge1} and {\tt edge2} from
the history file and converts them to the localized edges by using
the partitioning vector. This avoids the communication cost to
exchange each process's edges and the computation cost to choose the
edges to be assigned.
The disadvantage of the history file is that it cannot be used
if the program is run on a different number of processes from
when the file was created, because the edges and nodes being assigned to
each process dynamically change among different numbers of processes.
One efficient use of the history file is to create
it in advance for the various numbers of processes of interest.
As long as the user runs the application with any of those 
 numbers of processes,
an appropriate history can be chosen to reduce communication and
computation costs.
If there is no history file,
the edges in each process are distributed
by reading all the data in parallel and performing 
a ring-oriented communication.

If at least a node of an edge has been partitioned to a process, the edge is
assigned to the process. For example, edge {\tt 0} is assigned both to
process 0 and 1 because one node of the edge, {\tt edge1 0}, has been
partitioned to process 0 and the other node, {\tt edge2 1},
has been partitioned
to process 1. This edge is a ghost edge of both processes
being stored to minimize communication volumes.

For storing the partitioned edges and nodes, including the ghost ones,
a certain amount of memory space is initially allocated to each process.
When the entire memory space is occupied by the partitioned data,
it is automatically doubled by adjusting the memory size.
This prevents the system from looking through the entire data
in two steps, one step to decide the size of memory space and
the other step to actually store the data in the memory space.

After the edges and nodes are distributed, the edges in each process are
moved to the next process located at a ring network.
In the example, process 0 receives edges {\tt 2} and {\tt 3}, and
process 1 receives edges {\tt 0} and {\tt 1} to partition them
 as described above.
After finishing the edge distribution, edges {\tt 0} and {\tt 2} are
assigned to process 0, and edges {\tt 0}, {\tt 1}, and {\tt 3} are
assigned to process 1. Similarly, nodes {\tt 0}, {\tt 1}, and {\tt 3} are
assigned to process 0, and nodes {\tt 0}, {\tt 1}, {\tt 2}, and {\tt 4} are
assigned to process 1.
In Figure~\ref{fig:API2}, {\tt partitioned\_edge} contains the edges
assigned to each process, and {\tt vector} contains the nodes assigned to it.
These are the two
map arrays to distribute the physical data associated with each edge
and node, respectively.

If the {\em SDM\_index\_registry} was
executed for the first time and no history file was created earlier,
the metadata of the partitioned edges,
such as the partitioned size of each process, is stored in the
database tables {\em index\_table} and {\em index\_history\_table}.
Also, the partitioned edges are asynchronously written to a history file
to be retrieved in subsequent runs requiring the same edge distribution.
The use of the {\em SDM\_index\_registry} is optional.
If the user does not call the {\em SDM\_index\_registry},
no history file is created
after partitioning the edges.

In order to import and partition data {\tt x} and {\tt y} in
the {\em SDM\_import}, the {\em SDM\_data\_view} must be called to
define the data mapping between a noncontiguous global view of the file
and a local view of the processor memory. Using the data mapping,
 in the {\em SDM\_import}, the associated data is irregularly distributed
by calling a collective MPI-IO function. 
In the {\em SDM\_release\_importlist},
the structures being used to import data
in the file handle are free.

Figure~\ref{fig:API1} shows the steps to write two data sets, {\tt p} and
{\tt q}, after completing the computations at each checkpoint.
Before writing {\tt p} and {\tt q}, the data mapping
to write is defined in the {\em SDM\_data\_view}
using the map array ({\tt vector}) associated with
the node partition.

SDM supports three different ways of organizing data in files.  In
level 1, each data set generated at each time step
is written to a separate file.
This file organization is simple, but it
incurs the cost of a file-open, file-view to define the visible portion of
a file for each process and a file-close at each time step.
In level 2, each data set (within a group) is written to a separate
file, but different iterations of the same data set are appended to
the same file. This method results in a smaller number of files
and smaller file-open and file-view costs.
The offset in the file where data is appended is stored in the
{\em execution\_table}.
In level 3, all iterations of all data sets belonging to a group are
stored in a single file. As in
level~2, the file offset for each data set is stored in the
{\em execution\_table} by process~0 in the {\em SDM\_write} function.
The idea is that if a file system has high file-open and file-close costs, and
an application generates a high file-view cost, as in irregular
applications, SDM can generate a very small number of files.
However, if an application produces
a large number of data sets with a large problem size, level 3 file
organization would result in very large files, which may degrade
the performance.

\begin{figure}[h]
\centerline{\psfig{figure=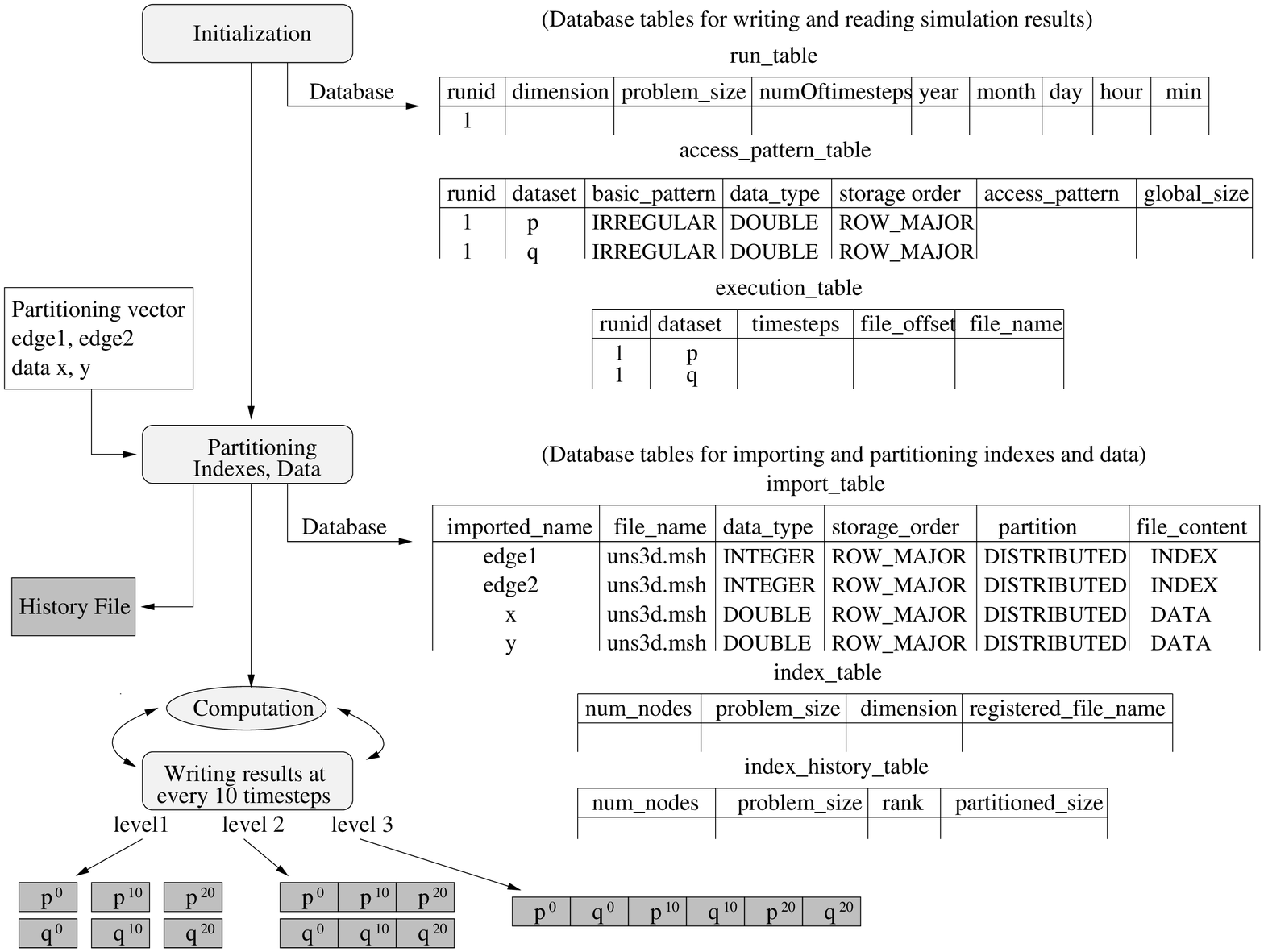,height=3.5in,width=3.2in}}
\caption{SDM execution flow to solve for the example in Figure~\ref{fig:problem}}
\label{fig:flow}
\end{figure}

Figure~\ref{fig:flow} depicts the metadata storage in the database and
the organization of data in files in SDM for the example 
in Figure~\ref{fig:problem}.

\Section{Performance Results \label{sec:perf}}

We obtained performance results on the SGI Origin2000 at Argonne
National Laboratory.  The Origin2000 has 128 processors and
10 Fibre Channel controllers connected to a total of 110 disks of
9 GBytes capacity each.  The file system on the Origin2000 is SGI's
XFS~\cite{XFS:Next,XFS:sweeney}. For the results, we used XFS buffered I/O
and MySQL~\cite{mysql-manual} to store the metadata.

The first application template that we benchmarked was a tetrahedral
vertex-centered unstructured grid code developed by W. K. Anderson
of the NASA Langley Research Center~\cite{fun3d}. This application
uses a partitioning vector generated from MeTis to partition the nodes
and edges in a mesh.
To evaluate SDM ported to the application, we used
about 18M edges and 2M nodes.
At the initial stage, the application
imports edges, four data arrays associated with edges, and another four
data arrays associated with nodes. The total imported data size was
about 807 MBytes.
As a result of computations, the application wrote 
about 21 MBytes of four data sets each
and 105 MBytes of a single data set.
Using 64 processors, we iterated the application template
 two time steps; at each time step,
five data sets were written to files.

The second application template that we ran was a Rayleigh-Taylor instability
application~\cite{lori:mesh} that is motivated by a joint project between the
University of Chicago and Argonne to study thermonuclear
flashes on astrophysical objects.
Whenever the current time reaches a certain point,
the application writes two data sets:
a single node data set associated with vertices in a mesh,
and a triangle data set associated with triangles on tetrahedral faces.
In the application template, we wrote about 36 MBytes of the node data set
and about 74 MBytes of the triangle data set at each time step.
Since we iterated the template five times, the total data size written
was approximately 550 MBytes.

\SubSection{Results for FUN3D}

\begin{figure}[h]
\centerline{\psfig{figure=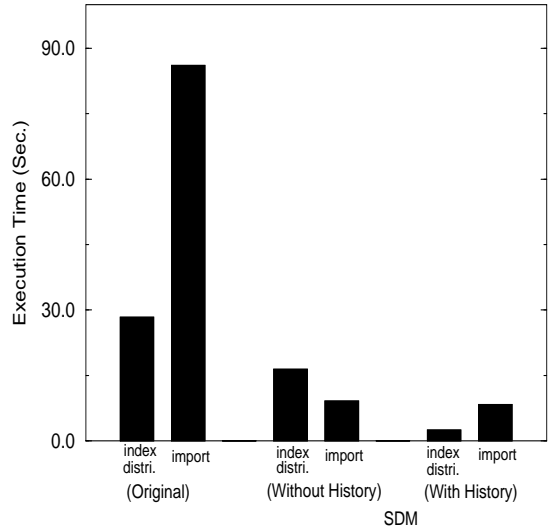,height=3in,width=3.2in}}
\caption{Execution time for partitioning indices and data in FUN3D}
\label{fig:import}
\end{figure}

Figure~\ref{fig:import} shows the bandwidth to import and partition
18M edges, four data sets each of 144 MBytes of data 
associated with edges, and
another four data sets each of 21 MBytes of data associated with nodes.
The original version of the application---without using SDM---performs all 
the I/O operations
 by a single process (process 0), which then broadcasts
data to other processes. SDM performs I/O in parallel from
all processes using MPI-IO.
The bar labeled {\tt index distri.} in Figure~\ref{fig:import}
shows the communication and computation costs
to partition the edges after importing them to the application.
Also, the bar labeled {\tt import} shows the cost of
reading the edges and eight data arrays.

The original application reads the edges in two steps: one
step to determine the amount of memory to
store the partitioned edges and the other step to actually
read the edges. SDM, however, extends the allocated memory
dynamically as needed (using C function {\tt realloc}) and
is therefore able to read the partitioned edges in a single step.
This contributes to the reduced cost of {\tt index distri.} when using
SDM.
When partitioning the edges with a history file,
the cost of {\tt index distri.}
is nothing but reading the history file of the edges in a contiguous way,
including the database cost to access the metadata.
Since the history file contains the already partitioned edges, there is
no need to import the edges; hence, the read cost
 in {\tt import} is reduced.

\begin{figure}[h]
\centerline{\psfig{figure=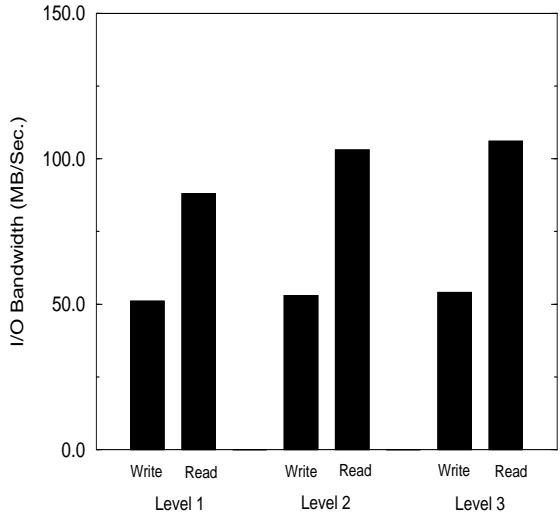,height=3in,width=3.2in}}
\caption{I/O bandwidth for reading and writing data in FUN3D}
\label{fig:app1}
\end{figure}

Figure~\ref{fig:app1} shows the I/O bandwidth for writing and then
reading back the data generated from the application using 64
processors.  The total data size was approximately 379 MBytes.  In
level 1, each data array is written to separate files, resulting in
the creation of 10 different files.  Each time the data array is
written to files, level 1 requires the cost for opening a file and
defining an MPI-IO {\em file view} to access the data from the portion
of the file pointed by the global file offset.  In level 2, however,
each data array generated at each time step is appended in five files,
generating five file-open and file-view costs. This reduced number of
files improves the I/O performance slightly.  In level
3, only two files are generated, resulting in the best I/O performance
among the three file organizations.  On the SGI Origin2000, the
difference between three file organizations is not significant because
the file-open cost is small.

\SubSection{Results of RT Application}

\begin{figure}[h]
\centerline{\psfig{figure=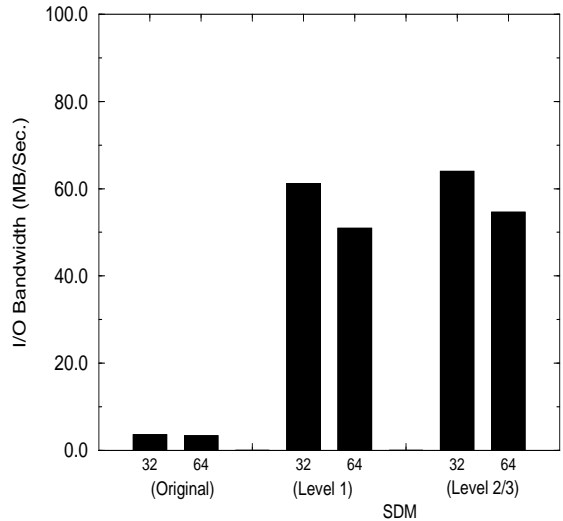,height=3in,width=3.2in}}
\caption{I/O bandwidth for RT}
\label{fig:app2}
\end{figure}

Figure~\ref{fig:app2} shows the I/O bandwidth for writing approximately
550 MBytes of data. In the original application, the write operation
is performed sequentially. In other words, after seeking
the starting position in a file, processes write their local portion
of data one by one.
When we ported the application to SDM, the I/O performance increased 
significantly because of the I/O optimizations of MPI-IO.

In SDM, we wrote the node data set according to the global node number of the
partitioned nodes, and wrote the triangle data set contiguously.
Since two data sets are written to files separately, SDM supports
two different ways of file organization: 
level 1 and level 2/3 (levels 2 and 3 are identical in this case).
As can be seen in Figure~\ref{fig:app2}, on the SGI Origin2000, changing the
file organization does not affect the I/O performance, since the cost of
file-open and file-view is very low.

When the number of processors increases to write the same
data size, we can see the degradation of the I/O performance.
With 32 processors, the data size being written at each time step is
about 1 MByte for the node data set and 2 MBytes for the triangle data set.
If the number of processors goes up to 64, the buffer size
of each process becomes smaller, resulting in the performance reduction.
Clearly, there is an optimal buffer size that shows the best I/O
performance.

\Section{Related Work \label{sec:related}}

Several efforts have sought to optimize I/O in parallel file
systems and runtime
libraries~\cite{benn94a,bord93b,corb96b,hube95a,kotz97a,madh96a,nieu97a,seam95b,thak96f}.
SRB (Storage Resource Broker)~\cite{baru98a} provides an uniform
interface to access various storage systems, such as file systems,
Unitree, HPSS and database objects. However, it does not fully support
the optimizations implemented in MPI-IO.  Shoshani et
al.~\cite{shos98a,shos99a} describe an architecture for optimizing
access to large volumes of scientific data stored on tapes.  The
Active Data Repository~\cite{kurc99a} and DataCutter~\cite{datacutter}
optimize storage, retrieval, and processing of very large
multidimensional datasets.  The main difference between our work and
other efforts in I/O is that SDM aims to combine the good
features of parallel file I/O and databases, whereas other efforts
focus on either parallel I/O or data management, not both.

\Section{Summary \label{sec:conc}}

We have described the SDM system, API, and implementation for I/O in
irregular applications. SDM provides an easy-to-use user interface for
managing large data sets and internally uses MPI-IO for
high-performance I/O and a database for storing metadata. We studied
the performance of SDM using two irregular applications: FUN3D and
RT. When we ported both applications to use SDM, there was a
significant improvement in I/O performance compared with the original
application. Also, we observed that using a history file for the
index distribution helped to reduce the computation and
communication costs.  However, changing the SDM file organization from
level 1 to level 3 did not greatly affect the performance on the SGI
Origin2000, because of its low file-open and file-view costs.

We plan to develop SDM further to support visualization applications
and to investigate whether SDM can effectively be used as a strategy
for implementing libraries such as HDF~\cite{hdf94a} and
netCDF~\cite{netc91a}.

%------------------------------------------------------------------------- 
\bibliographystyle{latex8}
\bibliography{bib1,bib2,bib3,bib4,bib5}
%\bibliography{/homes/jano/BIBTEX/pario-beta,/homes/jano/BIBTEX/main,/homes/jano/BIBTEX/MORE,/homes/jano/BIBTEX/SC,/homes/thakur/tex/bib/papers}
%\bibliography{latex8}

\end{document}